\documentstyle[12pt]{article}
\begin{document}

\title
{{\bf Dynamics of vortex and magnetic lines in ideal hydrodynamics and MHD}}
\author{E.A.Kuznetsov\footnote{E-mail: kuznetso@itp.ac.ru} \  and 
V.P.Ruban\footnote{E-mail: ruban@itp.ac.ru} \\
{\it Landau Institute for Theoretical Physics}\\
{\it 2 Kosygin str., 117334 Moscow, Russia}}
\date{}
\maketitle

\begin{abstract}
 Vortex line and magnetic line
representations are introduced for description of flows in
ideal hydrodynamics and MHD, respectively.
For incompressible fluids it is shown that
the equations of motion for vorticity ${\bf \Omega}$ and
magnetic field with the help of this transformation 
follow from the variational
principle. By means of this representation it is possible
to integrate  the system of hydrodynamic type with
the Hamiltonian ${\cal{H}}=\int |{\bf \Omega}| d{\bf r}$.
 It is also demonstrated that these representations allow to
remove from the noncanonical Poisson
brackets, defined on the space of divergence-free vector
fields, degeneracy connected with the vorticity frozenness for the Euler
equation  and with magnetic field frozenness for ideal MHD.
For MHD a new Weber type transformation is found. It is shown
how this transformation can be obtained from the two-fluid model
when electrons and ions can be considered as two independent fluids.
The Weber type transformation for ideal MHD gives the whole Lagrangian
vector invariant. When this invariant is absent this transformation
coincides with the Clebsch representation analog introduced by
Zakharov and Kuznetsov.
\end{abstract}

\newpage
\setcounter{equation}{0}
\section{Introduction}

There are a large number of works devoted to the Hamiltonian description of
the ideal hydrodynamics (see, for instance, the review \cite{ZK} and the
references therein). This question was first studied by Clebsch (a citation
can be found in Ref. \cite{lamb}), who introduced for nonpotential flows of
incompressible fluids a pair of variables $\lambda $ and $\mu $ (which later
were called as the Clebsch variables). A fluid dynamics in these variables
is such that vortex lines represent themselves intersection of surfaces $%
\lambda =\mbox{const}$ and $\mu =\mbox{const}$ and these quantities, being
canonical conjugated variables, remain constant by fluid advection.
However, these
variables, as known (see, i.e.,\cite{KM}) describe only partial type of
flows. If $\lambda $ and $\mu $ are single-valued functions of coordinates
then the linking degree of vortex lines characterizing by the Hopf invariant
\cite{fr} occurs to be equal to zero. For arbitrary flows the Hamiltonian
formulation of the equation for incompressible ideal hydrodynamics was given
by V.I.Arnold \cite{arnold1,arnold2}. The Euler equations for the velocity
curl ${\bf \Omega }=\mbox{curl}\,{\bf v}$
\begin{equation}
\frac{\partial {\bf \Omega }}{\partial t}=\mbox{curl}\,[{\bf v\times \Omega }%
],\,\,\,\mbox{div}\,{\bf v}=0  \label{euler}
\end{equation}
are written in the Hamiltonian form,
\begin{equation}
\frac{\partial {\bf \Omega }}{\partial t}=\{{\bf \Omega },{\cal {H}\},}
\label{euler1}
\end{equation}
by means of the noncanonical Poisson brackets \cite{KM}
\begin{equation}
\{F,G\}=\int \left( {\bf \Omega }\left[ \mbox{curl}\,\frac{\delta F}{\delta
{\bf \Omega }}\times \mbox{curl}\,\frac{\delta G}{\delta {\bf \Omega }}%
\right] \right) {d{\bf r}}  \label{bracket}
\end{equation}
where the Hamiltonian
\begin{equation}
{\cal H}_h=-\frac 12\int {\bf \Omega }\Delta ^{-1}{\bf \Omega }d^3{\bf r},
\label{Hhydro}
\end{equation}
coincides with the total fluid energy.

In spite of the fact that the bracket (\ref{bracket}) allows to describe
flows with arbitrary topology its main lack is a degeneracy. By this reason
it is impossible to formulate the variational principle on the whole space $%
{\cal S}$ of divergence-free vector fields.

The cause of the degeneracy, namely, presence of Casimirs annulling the
Poisson bracket, is connected with existence of the special symmetry formed
the whole group - the relabeling group of Lagrangian markers (for details
see the reviews \cite{salmon,ZK}). All known theorems about the vorticity
conservation (the Ertel's, Cauchy's and Kelvin's theorems, the frozenness of
vorticity and conservation of the topological Hopf invariant) are a sequence
of this symmetry. The main of them is the frozenness of vortex lines into
fluid. This is related to the local Lagrangian invariant -- the Cauchy
invariant. The physical meaning of this invariant consists in that any fluid
particle remains all the time on its own vortex line.

The similar situation takes place also for ideal magneto-hydrodynamics (MHD)
for barotropic fluids:
\begin{equation}
\rho _t+\nabla (\rho {\bf v})=0,  \label{rho-t}
\end{equation}
\begin{equation}
{\bf v}_t+({\bf v}\nabla ){\bf v}=-\nabla w(\rho )+\frac 1{4\pi \rho }[%
\mbox{curl}\,{\bf h}\times {\bf h}],  \label{v-t}
\end{equation}
\begin{equation}
{\bf h}_t=\mbox{curl}[{\bf v}\times {\bf h}].  \label{h-t}
\end{equation}
Here $\rho $ is a plasma density, $w(\rho)$\ plasma entalpy, ${\bf v}$ and
${\bf h}$ are velocity and magnetic fields, respectively. As well known
(see, for instance, \cite{landau}-\cite{Kuvshinov}),
the MHD equations possesses one
important feature -- frozenness of magnetic field into plasma which is
destroyed only due to dissipation (by finite conductivity). For ideal MHD
combination of the continuity equation (\ref{rho-t}) and the induction
equation (\ref{h-t}) gives the analog of the Cauchy invariant for MHD.

The MHD equations of motion (\ref{rho-t}-\ref{h-t}) can be also represented
in the Hamiltonian form,
\begin{equation}
\rho _t=\{\rho ,{\cal H}\}\qquad {\bf h}_t=\{{\bf h,}{\cal H}\},\qquad {\bf v%
}_t=\{{\bf v,}{\cal H\}},  \label{dp/dt}
\end{equation}
by means of the noncanonical Poisson brackets \cite{MG}:
\begin{equation} \label{compres_bracket}
\{F,G\}=\int \left( \frac{{\bf h}}\rho \cdot \left( \left[ \mbox{curl}\frac{%
\delta F}{\delta {\bf h}}\times \frac{\delta G}{\delta {\bf v}}\right]
-\left[ \mbox{curl}\frac{\delta G}{\delta {\bf h}}\times \frac{\delta F}{%
\delta {\bf v}}\right] \right) \right) d^3{\bf r}+
\end{equation}
\[
+\int \left( \frac{\mbox{curl}\ {\bf v}}\rho \cdot \left[ \frac{\delta F}{%
\delta {\bf v}}\times \frac{\delta G}{\delta {\bf v}}\right] \right) d^3{\bf %
r}+\int \left( \frac{\delta G}{\delta \rho }\nabla \left( \frac{\delta F}{%
\delta {\bf v}}\right) -\frac{\delta F}{\delta \rho }\nabla \left( \frac{%
\delta G}{\delta {\bf v}}\right) \right) d^3{\bf r}.
\]
This bracket is also degenerated. For instance, the integral $\int {\bf (v,h)%
}d{\bf r}$, which characterizes mutual linkage knottiness of vortex and
magnetic lines, is one of the Casimirs for this bracket.

The analog of the Clebsch representation in MHD serves a change of variables
suggested in 1970 by Zakharov and Kuznetsov \cite{ZK70}:
\begin{equation}
{\bf v}=\nabla \phi +\frac{\lbrack {\bf h}\times \mbox{curl}\ {\bf S}\rbrack
}{\rho }.  \label{phiHS}
\end{equation}
New variables $(\phi ,\rho )$ and ${\bf h,S}$ represent two pairs
canonically conjugated quantities with the Hamiltonian coinciding with the
total energy
\[
{\cal H}=\int \left( \rho \frac{{\bf v}^{2}}{2}+\rho \varepsilon (\rho )+%
\frac{{\bf h}^{2}}{8\pi }\right) d{\bf r}.
\]

In the present paper we suggest a new approach of the degeneracy resolution
of the noncanonical Poisson brackets by introducing new variables, i.e.,
Lagrangian markers labeling each vortex lines for ideal hydrodynamics or
magnetic lines in the MHD case.

The basis of this approach is the integral representation for the
corresponding frozen-in field, namely, the velocity curl for the Euler
equation and magnetic field for MHD. We introduce new objects, i.e., the
vortex lines or magnetic lines and obtain the equations of motion for them.
This description is a mixed Lagrangian-Eulerian description, when each
vortex (or magnetic) line is enumerated by Lagrangian marker, but motion
along the line is described in terms of the Eulerian variables. Such
representation removes all degeneracy from the Poisson brackets connected
with the frozenness, remaining the equations of motion to be gauge
invariant with respect to re-parametrization of each line. Important, that
the equations for line motion, as the equations for curve deformation, are
transverse to the line tangent.

It is interesting that the line representation also solves another problem -
the equations of line motion follow from the variational principle, being
Hamiltonian.

This approach allows also simply enough to consider the limit of narrow
vortex (or magnetic) lines. For two-dimensional flows in hydrodynamics this
''new'' description corresponds to the well-known fact, namely, to the
canonical conjugation of $x$ and $y$ coordinates of vortices
(see, for instance, \cite{lamb}).

The Hamiltonian structure introduced makes it possible to integrate
the three-di\-men\-sional Euler equation (\ref{euler1}) with Hamiltonian 
${\cal H}=\int |{\bf \Omega }|d{\bf r}$. 
In terms of the vortex lines the given
Hamiltonian is decomposed into a set of Hamiltonians of noninteracting
vortex lines. The dynamics of each vortex lines is, in turn, described by
the equation of a vortex induction which can be reduced by the Hasimoto
transformation \cite{Hasimoto} to the integrable one-dimensional nonlinear
Schrodinger equation.

For ideal MHD a new representation - analog of the Weber transformation - is
found. This representation contains the whole vector Lagrangian invariant.
In the case of ideal hydrodynamics this invariant provides conservation of
the Cauchy invariant and, as a sequence, all known conservation laws for
vorticity (for details see the review \cite{ZK}). It is important that all
these conservation laws can be expressed in terms of observable variables.
Unlike the Euler equation, these vector Lagrangian invariants for the MHD
case can not be expressed in terms of density, velocity and magnetic field.
It is necessary to tell that the analog of the Weber transformation for MHD
includes the change of variables (\ref{phiHS}) as a partial case. The
presence of these Lagrangian invariants in the transform provides
topologically nontrivial MHD flows.

The Weber transform and its analog for MHD play a key role in constructing
the vortex line (or magnetic line) representation. This representation is
based on the property of frozenness. Just therefore by means of such
transform the noncanonical Poisson brackets become non-degenerated in these
variables and, as a result, the variational principle may be formulated.
Another peculiarity of this representation is its locality, establishing the
correspondence between vortex (or magnetic) line and vorticity (or magnetic
field). This is a specific mapping, mixed Lagrangian-Eulerian, for which
Jacobian of the mapping can not be equal to unity for incompressible fluids
as it is for pure Lagrangian description.

\setcounter{equation}{0}
\section{General remarks}

We start our consideration from some well known facts, namely, from the
Lagrangian description of the ideal hydrodynamics.

In the Eulerian description for barotropic fluids, pressure
$p=p(\rho )$, we have
coupled equations - discontinuity equation for density $\rho $ and the Euler
equation for velocity:
\begin{equation}
\rho _{t}+\mbox{div}\,\rho {\bf v}=0,  \label{cont}
\end{equation}
\begin{equation}
{\bf v}_{t}+{\bf (v\nabla )v}=-\nabla w(\rho ),\qquad dw(\rho )=dp/\rho .
\label{velocity}
\end{equation}
In the Lagrangian description each fluid particle has its own label. This is
three-dimen\-sional vector ${\bf a}$, so that particle position at time $t$ is
given by the function
\begin{equation}
{\bf x=x(a,}t).  \label{map}
\end{equation}
Usually initial position of particle serves the Lagrangian marker: ${\bf %
a=x(a,}0)$.

In the Lagrangian description the Euler equation (\ref{velocity}) is nothing
more than the Newton equation:
\[
\ddot{{\bf x}}=-\nabla w.
\]
In this equation the second derivative with respect to time $t$ is taken for
fixed ${\bf a}$, but the r.h.s. of the equation is a function of $t$ and $%
{\bf x}$. Excluding from the latter the $x$-dependence, the Euler equation
takes the form:
\begin{equation}
\ddot{x}_{i}\frac{\partial x_{i}}{\partial a_{k}}=-\frac{\partial w(\rho )}{%
\partial a_{k}}\,,  \label{Lag}
\end{equation}
where now all quantities are functions of $t$ and ${\bf a.}$

In the Lagrangian description the continuity equation (\ref{cont}) is easily
integrated and the density is given through the Jacobian of the mapping (\ref
{map}) $J=\mbox{det}(\partial x_{i}/\partial a_{k})$:
\begin{equation}
\rho =\frac{\rho _{0}({\bf a})}{J}.  \label{density}
\end{equation}

Now let us introduce a new vector,
\begin{equation}
u_k=\frac{\partial x_i}{\partial a_k}v_i\,,  \label{u}
\end{equation}
which has a meaning of velocity in a new curvilinear system of coordinates
or it is possible to say that this formula defines the transformation law
for velocity components. It is worth noting that (\ref{u}) gives the
transform for the velocity ${\bf v}$ as a {\it co-vector}.

The straightforward calculation gives that the vector ${\bf u}$ satisfies
the equation
\begin{equation}
\frac{du_k}{dt}=\frac \partial {\partial a_k}\left( \frac{{\bf v}^2}%
2-w\right) .  \label{motion}
\end{equation}
In this equation the right-hand-side represents gradient relative to ${\bf a}
$ and therefore the ''transverse'' part of the vector ${\bf u}$ will
conserve in time. And this gives the Cauchy invariant:
\begin{equation}
\frac d{dt}\mbox{curl }_a\,{\bf u}=0,  \label{curl}
\end{equation}
or
\begin{equation}
\mbox{curl}_a\,{\bf u=I}.  \label{Cauchy}
\end{equation}
If Lagrangian markers ${\bf a}$ are initial positions of fluid particles
then the Cauchy invariant coincides with the initial vorticity: ${\bf %
I=\Omega _0(a)}$. This invariant is expressed through instantaneous value of $%
{\bf \Omega (x},t)$ by the relation
\begin{equation}
{\bf \Omega _0(a)=}J{\bf (\Omega (x},t)\nabla ){\bf a(x},t)  \label{Ca}
\end{equation}
where ${\bf a=a(x,}t)$ is inverse mapping to (\ref{map}). Following from (%
\ref{Ca}) relation for ${\bf B=\Omega }/\rho ,$%
\[
B_{0i}(a)=\frac{\partial a_i}{\partial x_k}B_k(x,t),
\]
shows that, unlike velocity, ${\bf B}$ transforms as a vector.

By integrating the equation (\ref{motion}) over time $t$ we arrive at the
so-called Weber transformation
\begin{equation}
{\bf u(a},t)={\bf u}_{0}({\bf a})+\nabla _{a}\Phi ,  \label{weber}
\end{equation}
where the potential $\Phi $ obeys the Bernoulli equation:
\begin{equation}
\frac{d\Phi }{dt}=\frac{{\bf v}^{2}}{2}-w(\rho )  \label{Phi}
\end{equation}
with the initial condition: $\Phi |_{{t=0}}=0$. For such choice of $\Phi $ a
new function ${\bf u_{0}(a)}$ is connected with the ''transverse'' part of $%
{\bf u}$ by the evident relation
\[
\mbox{curl}_{a}\,{\bf u}_{0}({\bf a)=I}.
\]

The Cauchy invariant ${\bf I}$
characterizes the vorticity frozenness into fluid. It
can be got by standard way considering two equations - the equation for the
quantity ${\bf B=\Omega /\rho }$,
\begin{equation}
\frac{d{\bf B}}{dt}=({\bf B\nabla )v},  \label{froz}
\end{equation}
and the equation for the vector $\delta {\bf x=x(a+\delta a)-x(a)}$ between
two adjacent fluid particles:
\begin{equation}
\frac{d\delta {\bf x}}{dt}=({\delta {\bf x\nabla )v}},  \label{delta}
\end{equation}
The comparison of these two equations shows that if initially the vectors ${%
\delta {\bf x}}$ are parallel to the vector ${\bf B}$, then they will be
parallel to each other all time. This is nothing more than the statement of
the vorticity frozenness into fluid. Each fluid particle remains all the
time at its own vortex line. The combination of Eqs. (\ref{froz}) and (\ref
{delta}) leads to the Cauchy invariant. To establish this fact it is enough
to write down the equation for the Jacoby matrix $J_{ij}=\partial
x_{i}/\partial a_{j}$ which directly follows from (\ref{delta}):
\[
\frac{d}{dt}\ \frac{{\partial a_{i}}}{{\partial x}_{k}}=-\frac{\partial a_{i}%
}{\partial x_{j}}\frac{\partial v_{j}}{\partial x_{k}},
\]
that in combination with Eq. (\ref{froz}) gives conservation of the Cauchy
invariant (\ref{Cauchy}).

If now one comes back to the velocity field ${\bf v}$
then by use of Eqs. (\ref{u}) and (\ref{weber}) one can get that
\begin{equation}
{\bf v}=u_{0k}\nabla a_{k}+\nabla \Phi  \label{v}
\end{equation}
where gradient is taken with respect to ${\bf x}$. Here the equation for
potential $\Phi $ has the standard form of the Bernoulli equation:
\[
\Phi _{t}+({\bf v}\nabla )\Phi -\frac{{\bf v^{2}}}{2}+w(\rho )=0.
\]
It is interesting to note that relations (\ref{Cauchy}), as equations for
determination of ${\bf x(a,}t)$, unlike Eqs (\ref{motion}), are of the first
order with respect to time derivative. This fact also reflects in the
expression for velocity (\ref{v}) which can be considered as a result of the
partial integration of the equations of motion (\ref{motion}). Of course,
the velocity field given by (\ref{v}) contain two unknown functions: one is
the whole vector ${\bf a(x},t)$ and another is the potential $\Phi $.
For incompressible fluids the latter is determined from the condition $%
\mbox{div}\,{\bf v}=0$. In this case the Bernoulli equation serves for
determination of the pressure.

Another important moment connected with the Cauchy invariant is that it
follows from the variational principle (written in terms of Lagrangian
variables) as a sequence of relabelling symmetry remaining invariant the
action (for details, see the reviews \cite{salmon,ZK}). Passing from
Lagrangian to Hamiltonian in this description we have no any problems with
the Poisson bracket. It is given by standard way and does not contain any
degeneracy that the noncanonical Poisson brackets (\ref{bracket}) and (\ref
{compres_bracket}) have. One of the main purposes of this paper is to
construct such new description of the Euler equation (as well as the ideal
MHD) which, from one side, would allow to retain the Eulerian description,
as maximally as possible, but, from another side, would exclude from the
very beginning all remains from the gauge invariance of the complete Euler
description connected with the relabeling symmetry.

As for MHD, this system in one point has some common feature with the Euler
equation: it also possesses the frozenness property. The equation for ${\bf h}%
/\rho $ coincides with (\ref{froz}) and therefore dynamics of magnetic lines
is very familiar to that for vortex lines of the Euler equation. However,
this analogy cannot be continued so far because the equation of motion for
velocity differs from the Euler equation by the presence of pondermotive
force. This difference remains also for incompressible case.

\setcounter{equation} {0}
\section{Vortex line representation}

Consider the Hamiltonian dynamics of the divergence-free vector field ${\bf %
\Omega }({\bf r},t)$, given by the Poisson bracket (\ref{bracket}) with some
Hamiltonian ${\cal {\ H}}$ \footnote{{\normalsize The Hamiltonian (\ref
{Hhydro}) corresponds to ideal incompressible hydrodynamics.}}:
\begin{equation}
\frac{\partial {\bf \Omega }}{\partial t}=\mbox{curl}\left[ \mbox{curl}\,
\frac{\delta {\cal H}}{\delta {\bf \Omega }}\times {\bf \Omega }\right] .
\label{dOmega_dt}
\end{equation}

As we have said, the bracket (\ref{bracket}) is degenerate, as a result of
which it is impossible to formulate the variational principle on the entire
space ${\cal S}$ of solenoidal vector fields. It is known \cite{ZK} that
Casimirs $f$, annulling Poisson brackets, distinguish in ${\cal S}$
invariant manifolds ${\cal M}_{f}$ (symplectic leaves) on each of which it
is possible to introduce the standard Hamiltonian mechanics and accordingly
to write down a variational principle. We shall show that solution of this
problem for the equations (\ref{dOmega_dt}) is possible on the base of the
property of frozenness of the field ${\bf \Omega (r},t)$, which allows to
resolve all constrains, stipulated by the Casimirs, and gives the necessary
formulation of the variational principle.

To each Hamiltonian ${\cal {H}}$ - functional of ${\bf \Omega (r},t)$ - we
associate the generalized velocity
\begin{equation}
{\bf v}({\bf r})=\mbox{curl}\,\frac{\delta {\cal H}}{\delta {\bf \Omega }}.
\label{Vgeneral}
\end{equation}
However one should note that the generalized ${\bf v}({\bf r})$ is defined
up to addition of the vector parallel to ${\bf \Omega }$:
\[
\mbox{curl}\,\frac{\delta {\cal H}}{\delta {\bf \Omega }}\rightarrow %
\mbox{curl}\,\frac{\delta {\cal H}}{\delta {\bf \Omega }}+\alpha {\bf \Omega
},
\]
that in no way does change the equation for ${\bf \Omega }$. Under the
condition $({\bf \Omega }\cdot \nabla \alpha )=0$ a new generalized velocity
will have zero divergence and the frozenness equation (\ref{dOmega_dt}) can
be written already for the new ${\bf v}({\bf r})$. A gauge changing of the
generalized velocity corresponds to some addition of a Casimir \ to the
Hamiltonian :
\[
{\cal H}\rightarrow {\cal H}+f;\qquad \{f,..\}=0.
\]
Hence becomes clear that the transformation
\[
{\bf x=x}({\bf a},t)
\]
of the initial positions of fluid particles ${\bf x}({\bf a},0)={\bf a}$ by
the generalized velocity field ${\bf v}({\bf r})$ through solution of the
equation
\begin{equation}  \label{vel}
{\bf {\dot x} = v(x,}t)
\end{equation}
is defined ambiguously due to the ambiguous definition of ${\bf v}({\bf r})$
by means of (\ref{Vgeneral}). Therefore using full Lagrangian description to
the systems (\ref{dOmega_dt}) becomes ineffective.

Now we introduce the following general expression for ${\bf \Omega }({\bf r}%
) $, which is gauge invariant and fixes all topological properties of the
system that are determined by the initial field ${\bf \Omega }_{0}({\bf a})$%
\cite{KR}:
\begin{equation}
{\bf \Omega }({\bf r},t)=\int \delta({\bf r}-{\bf R}({\bf a},t))({\bf \Omega
}_{0}({\bf a})\nabla _{{\bf a}}){\bf R}({\bf a},t)d^{3}{\bf a}.
\label{OmegaR}
\end{equation}
Here now
\begin{equation}
{\bf r= R}({\bf a},t)  \label{particle}
\end{equation}
does not satisfy any more the equation (\ref{vel}) and, consequently,
the mapping Jacobian $J=\mbox{det}||\partial {\bf R}/\partial {\bf a}||$ is
not assumed to equal 1, as it was for full Lagrangian description of
incompressible fluids.

It is easily to check that from condition $(\nabla _{{\bf a}}{\bf \Omega }%
_{0}({\bf a}))=0$ it follows that divergence of (\ref{OmegaR}) is
identically equal to zero.

The gauge transformation
\begin{equation}
{\bf R}({\bf a})\rightarrow {\bf R}(\tilde{{\bf a}}_{\Omega _{0}}({\bf a}))
\label{kalibr}
\end{equation}
leaves this integral unchanged if $\tilde{{\bf a}}_{\Omega _{0}}$ is arisen
from ${\bf a}$ by means of arbitrary nonuniform translations along the field
line of ${\bf \Omega }_{0}({\bf a})$. Therefore the invariant manifold $%
{\cal M}_{\Omega _{0}}$ of the space ${\cal S}$, on which the variational
principle holds, is obtained from the space ${\cal R}:{\bf a}\rightarrow
{\bf R}$ of arbitrary continuous one-to-one three-dimensional mappings
identifying ${\cal R}$ elements that are obtained from one another with the
help of the gauge transformation (\ref{kalibr}) with a fixed solenoidal
field $\ {\bf \Omega }_{0}({\bf a})$.

The integral representation for ${\bf \Omega }$ (\ref{OmegaR}) is another
formulation of the frozenness condition - after integration of the relation (%
\ref{OmegaR}) over area $\sigma $, transverse to the lines of ${\bf \Omega }$%
, follows that the flux of this vector remains constant in time:
\[
\int_{\sigma (t)}({\bf \Omega },d{\bf S_{r}})=\int_{\sigma (0)}({\bf \Omega
_{0}},d{\bf S_{a}}).
\]
Here $\sigma (t)$ is the image of $\sigma (0)$ under the transformation (\ref
{particle}).

It is important also that ${\bf \Omega _0(a)}$ can be expressed explicitly
in terms of the instantaneous value of the vorticity and the mapping ${\bf %
a=a(r},t)$, inverse to (\ref{particle}). By integrating over the variables $%
{\bf a}$ in the relation (\ref{OmegaR}),
\begin{equation}
{\bf \Omega }({\bf R})=\frac{({\bf \Omega }_0({\bf a})\nabla _{{\bf a}}){\bf %
R}({\bf a})}{\mbox{det}||\partial {\bf R}/\partial {\bf a}||},  \label{O_det}
\end{equation}
where ${\bf \Omega _0(a)}$ can be represented in the form:
\begin{equation}
{\bf \Omega _0}({\bf a})=\mbox{det}||\partial {\bf R}/\partial {\bf a}||(%
{\bf \Omega }({\bf r})\nabla ){\bf a}.  \label{C}
\end{equation}
This formula is nothing more than the Cauchy invariant (\ref{Cauchy}). We
note that according to Eq. (\ref{O_det}) the vector ${\bf b=(\Omega }_0({\bf %
a})\nabla _{{\bf a}}){\bf R}({\bf a})$ is tangent to ${\bf \Omega }({\bf R})$%
. It is natural to introduce parameter $s$ as an arc length of the initial
vortex lines ${\bf \Omega _0}({\bf a})$ so that
\[
{\bf b}=\Omega _0(\nu )\frac{\partial {\bf R}}{\partial s}.
\]
In this expression $\Omega _0$ depends on the transverse parameter $\nu $
labeling each vortex line. In accordance with this, the representation (\ref
{OmegaR}) can be written in the form
\begin{equation}
{\bf \Omega }({\bf r},t)=\int \Omega _0(\nu )d^2\nu \int \delta ({\bf r}-%
{\bf R}(s,\nu ,t))\frac{\partial {\bf R}}{\partial s}ds,  \label{rho}
\end{equation}
whence the meaning of the new variables becomes clearer: To each vortex line
with index $\nu $ there is associated the closed curve
\[
{\bf r}={\bf R}(s,\nu ,t),
\]
and the integral (\ref{rho}) itself is a sum over vortex lines. We notice
that the parametrization by introduction of $s$ and $\nu $ is local.
Therefore as global the representation (\ref{rho}) can be used only for
distributions with closed vortex lines.

To get the equation of motion for ${\bf R(\nu ,}s,t)$ the representation (%
\ref{rho}) (in the general case - (\ref{OmegaR})) must be substituted in the
Euler equation (\ref{dOmega_dt}) and then a Fourier transform with respect to
spatial coordinates performed. As a result of simple integration one can
obtain:
\[
\left[ {\bf k}\times \int \Omega _{0}(\nu )d^{2}\nu \int dse^{-i{\bf kR}%
}\lbrack {\bf R}_{s}\times \{{\bf R}_{t}(\nu,s,t)-{\bf v(R},t)\}\rbrack
\right] =0.
\]
This equation can be resolved by putting integrand equal identically to
zero:
\begin{equation}
\lbrack {\bf R}_{s}\times {\bf R}_{t}(\nu ,s,t)\rbrack =\lbrack {\bf R}_{s}%
\times {\bf v(R},t)\rbrack .  \label{R}
\end{equation}
With this choice there remains the freedom in both changing the parameter $s$
and relabelling the transverse coordinates $\nu$. In the general case of
arbitrary topology of the field ${\bf \Omega _{0}(a)}$ the vector ${\bf R}_{s}$
in the equation (\ref{R}) must be replaced by the vector ${\bf b}=({\bf %
\Omega }_{0}({\bf a})\nabla _{{\bf a}}){\bf R}({\bf a},t)$. Notice that, as
it follows from (\ref{R}) and (\ref{O_det}), a motion of a point on the
manifold ${\cal M}_{\Omega _{0}}$ is determined only by the transverse to $%
{\bf \Omega }({\bf r})$ component of the generalized velocity.

The obtained equation (\ref{R}) is the equation of motion for vortex lines.
In accordance with (\ref{R}) the evolution of each vector ${\bf R}$ is
principally transverse to the vortex line. The longitudinal component of
velocity does not effect on the line dynamics.

The description of vortex lines with the help of equations (\ref{rho}) and (%
\ref{R}) is a mixed Lagrangian-Eulerian one: The parameter $\nu $ has a
clear Lagrangian origin whereas the coordinate $s$ remains Eulerian.

\setcounter{equation} {0}
\section{Variational principle}

The key observation for formulation of the variational principle is that the
following general equality holds for functionals that depend only on $%
{\bf \Omega }$:
\begin{equation}
\left[ {\bf b}\times \mbox{curl}\left( \frac{\delta F}{\delta {\bf \Omega }(%
{\bf R})}\right) \right] =\frac{\delta F}{\delta {\bf R}({\bf a})}\Big|_{%
{\bf \Omega }_{0}}.  \label{peresch}
\end{equation}
For this reason, the right-hand-side of (\ref{R}) equals the variational
derivative $\delta {\cal H}/\delta{\bf R}$:
\begin{equation}
\left[ ({\bf \Omega }_{0}({\bf a})\nabla _{{\bf a}}){\bf R}({\bf a})\times
{\bf R}_{t}({\bf a})\right] =\frac{\delta {\cal H}\{{\bf \Omega }\{{\bf R}%
\}\}}{\delta {\bf R}({\bf a})}\Big|_{{\bf \Omega }_{0}}.  \label{main}
\end{equation}
It is not difficult to check now that the equation (\ref{main}) described
dynamics of
vortex line is equivalent to the requirement of extremum of the action ($%
\delta S=0$) with the Lagrangian \cite{KR}
\begin{equation}
{\cal L}=\frac{1}{3}\int d^{3}{\bf a}(\lbrack {\bf R}_{t}({\bf a})\times
{\bf R}({\bf a})\rbrack \cdot ({\bf \Omega }_{0}({\bf a})\nabla _{{\bf a}})%
{\bf R}({\bf a}))-{\cal H}(\{{\bf \Omega }\{{\bf R}\}\}).  \label{LAGRANGIAN}
\end{equation}

Thus, we have introduced the variational principle for the Hamiltonian
dynamics of the divergence-free vector field topologically equivalent to $%
{\bf \Omega }_{0}({\bf a})$.

Let us discuss some properties of the equations of motion (\ref{main}),
which are associated with excess parametrization of elements of ${\cal M}%
_{\Omega _{0}}$ by objects from ${\cal R}$. We want to pay attention to the
fact that From Eq. (\ref{peresch}) follows the property that the vector $%
{\bf b}$ and ${\delta F}/{\delta {\bf R}({\bf a})}$ are orthogonal for all
functionals defined on ${\cal M}_{\Omega _{0}}$. In other words the
variational derivative of the gauge-invariant functionals should be
understood (specifically, in (\ref{peresch})) as
\[
\hat{P}\frac{\delta F}{\delta {\bf R}({\bf a})},
\]
where $\hat{P}_{ij}=\delta _{ij}-\tau _{i}\tau _{j}$ is a projector and $%
{\bf \tau }={\bf b/|b|}$ a unit tangent (to vortex line) vector. Using this
property as well as the transformation formula (\ref{peresch}) it is
possible, by a direct calculation of the bracket (\ref{bracket}), to obtain
the Poisson bracket (between two gauge-invariant functionals) expressed in
terms of vortex lines:
\begin{equation}
\{F,G\}=\int \frac{d^{3}{\bf a}}{|{\bf b}|^{2}}\left( {\bf b}\cdot \left[
\hat{P}\frac{\delta F}{\delta {\bf R}({\bf a})}\times \hat{P}\frac{\delta G}{%
\delta {\bf R}({\bf a})}\right] \right) .  \label{NonCanR}
\end{equation}

The new bracket (\ref{NonCanR}) does not contain variational derivatives
with respect to ${\bf \Omega }_{0}({\bf a})$. Therefore, with respect to the
initial bracket the Cauchy invariant ${\bf \Omega }_{0}({\bf a})$ is a
Casimir fixing the invariant manifolds ${\cal M}_{\Omega _{0}}$ on which it
is possible to introduce the variational principle (\ref{LAGRANGIAN}).

In the case of the hydrodynamics of a superfluid liquid a Lagrangian of the
form (\ref{LAGRANGIAN}) was apparently first used by Rasetti and Regge \cite
{RR} to derive an equation of motion, identical to Eq. (\ref{R}), but for a
separate vortex filament. Later, on the base of the results \cite{RR},
Volovik and Dotsenko Jr. \cite{VD} obtained the Poisson bracket between the
coordinates of the vortices and the velocity components for a continuous
distribution of vortices. The expression for these brackets can be
extracted without difficulty from the general form for the Poisson brackets
(\ref{NonCanR}) . However, the noncanonical Poisson brackets obtained in
\cite{RR,VD} must be used with care. Their direct application gives for the
equation of motion of the coordinate of a vortex filament an answer that is
not gauge-invariant. For a general variation, which depends on time,
additional terms describing flow along a vortex appear in the equation of
motion. For this reason, the dynamics of curves (including vortex lines) is
in principle "transverse" with respect to the curve itself.

We note that for two-dimensional (in the $x-y$ plane) flows the variational
principle for action with the Lagrangian (\ref{LAGRANGIAN}) leads to the
well-known fact that $X(\nu ,t)$- and $Y(\nu ,t)$- coordinates of each
vortex are canonically conjugated quantities (see \cite{lamb}).

\setcounter{equation} {0}
\section{Integrable hydrodynamics}

Now we present an example of the equations of the hydrodynamic type (\ref
{dOmega_dt}), for which transition to the representation of vortex lines
permits to establish of the fact of their integrability \cite{KR}.

Consider the Hamiltonian
\begin{equation}
{\cal H}\{{\bf \Omega (r)}\}=\int |{\bf \Omega }|d{\bf r}  \label{hamilt}
\end{equation}
and the corresponding equation of frozenness (\ref{dOmega_dt}) with the
generalized velocity
\[
{\bf v}=\mbox{curl}\left( {\bf \Omega }/{\Omega }\right) .
\]
We assume that vortex lines are closed and apply the representation (\ref
{rho}). Then due to (\ref{O_det}) the Hamiltonian in terms of vortex lines
is decomposed as a sum of Hamiltonians of vortex lines:
\begin{equation}
{\cal H}\{{\bf R}\}=\int |\Omega _{0}(\nu )|d^{2}\nu \int \left| \frac{%
\partial {\bf R}}{\partial s}\right| ds.  \label{ham}
\end{equation}
The standing here integral over $s$ is the total length of a vortex line
with index $\nu $. According to (\ref{main}), with respect to these
variables the equation of motion for the vector ${\bf R}(\nu ,s)$ is local,
it does not contain terms describing interaction with other vortices:
\begin{equation}
\eta \lbrack {\bf \tau }\times {\bf R}_{t}(\nu ,s,t)\rbrack ={\bf \lbrack
\tau \times \lbrack \tau \times \tau} _{s}\rbrack \rbrack .  \label{motion1}
\end{equation}
Here $\eta =\mbox{sign}(\Omega _{0})$, $ {\bf \tau =R}_{s}/|{\bf R}_{s}|$
is the unit vector tangent to the vortex line.

This equation is invariant against changes $s\rightarrow \tilde{s}(s,t)$.
Therefore the equation (\ref{motion1}) can be resolved relative to ${\bf R}_t$
up to a shift along the vortex line -- the transformation unchanged the
vorticity ${\bf \Omega }$. This means that to find ${\bf \Omega }$ it is
enough to have one solution of the equation
\begin{equation}
\eta |{\bf R}_{s}|{\bf R}_{t}={\bf \lbrack \tau \times \tau}_s \rbrack
+\beta{\bf R}_{s},  \label{motion2}
\end{equation}
which follows from (\ref{motion1}) for some value of $\beta $. Arisen from
here equation for ${\bf \tau }$ as a function of filament length $l$ ($dl=|%
{\bf R}_{s}|ds$) and time $t$ (by choosing a new value $\beta =0$) reduces
to the integrable one-dimensional Landau-Lifshits equation for a Heisenberg
ferromagnet:
\[
\eta \frac{\partial {\bf \tau }}{\partial t}=\left[ {\bf \tau }\times \frac{%
\partial ^{2}{\bf \tau }}{\partial l^{2}}\right] .
\]

This equation is gauge-equivalent to the 1D nonlinear Schr\"{o}dinger
equation \cite{ZT} and, for instance, can be reduced to the NLSE by means of
the Hasimoto transformation \cite{Hasimoto}:
\[
\psi (l,t)=\kappa (l,t)\cdot \exp (i\int^{l}\chi (\tilde{l},t)d\tilde{l}),
\]
where $\kappa (l,t)$ is a curvature and $\chi (l,t)$ the line torsion.

The considered system with the Hamiltonian (\ref{hamilt}) has direct
relation to hydrodynamics. As known (see the paper \cite{Hasimoto} and
references therein), the local approximation for thin vortex filament (under
assumption of smallness of the filament width to the characteristic
longitudinal scale) leads to the Hamiltonian (\ref{ham}) but only for one
separate line. Respectively, the equation (\ref{dOmega_dt}) with the
Hamiltonian (\ref{hamilt}) can be used for description of motion of a few
number of vortex filaments, thickness of which is small compared with a
distance between them. In this case (nonlinear) dynamics of each filament is
independent upon neighbor behavior. In the framework of this model
singularity appearance ( intersection of vortices) is of an inertial
character very similar to the wave breaking in gas-dynamics. Of course, this
approximation does not work on distances between filaments comparable with
filament thickness.

It should be noted also that for the given approximation the Hamiltonian of
vortex line is proportional to the filament line whence its conservation
follows that, however, in no cases is adequate to behavior of vortex
filaments in turbulent flows where usually process of vortex filament
stretching takes place. It is desirable to have the better model free from
this lack. A new model must necessarily describe nonlocal effects.

In addition we would like to say that the list of equations (\ref{dOmega_dt}%
) which can be integrated with the help of representation (\ref{rho}) is not
exhausted by (\ref{hamilt}). So, the system with the Hamiltonian
\begin{equation}
{\cal H}_\chi \{{\bf \Omega (r)}\}=\int |{\bf \Omega }|\chi d{\bf r}
\label{hamilt_chi}
\end{equation}
is gauge equivalent to the modified KdV equation
\[
\psi _t+\psi _{lll}+\frac 32|\psi |^2\psi _l=0\,\,\,-
\]
the second one after NLSE in the hierarchy generated by Zakharov-Shabat
operator. As against previous model (\ref{hamilt}) some physical application
of (\ref{hamilt_chi}) has not yet been found.

\setcounter{equation}{0}
\section{Lagrangian description of MHD}

Consider now how the relabelling symmetry works in the ideal MHD.
First, rewrite
equations of motion (\ref{rho-t}-\ref{h-t}) in the Lagrangian
representation by introducing markers ${\bf a}$ for fluid particles
\[
{\bf x=x(a},t)
\]
with
$$
{\bf v(x},t)={\dot {{\bf x}}}{\bf (a},t).
$$
In this case the continuity
equation (\ref{rho-t})
and the equation for magnetic field (\ref{h-t}) can be integrated.
The density and
the magnetic field are expressed in terms of the Jacoby matrix by means of Eq.
(\ref{density}) and by the equation
\begin{equation}
B_i(x,t)=\frac{\partial x_i}{\partial a_k}B_{0k}(a)\,,  \label{htr}
\end{equation}
where ${\bf B=h}/\rho .$ In the latter transformation the Jacoby matrix
serves the evolution operator for vector ${\bf B.}$ The vector ${\bf B}$, in
turn, transforms as a vector.

In terms of Lagrangian variables the equation of motion (\ref{v-t}) is
written as follows
\begin{equation}
\frac{\partial x_i}{\partial a_k}\ddot {x_i}=-\frac{\partial w(\rho )}{%
\partial a_k}+\frac J{4\pi \rho _0({\bf a})}
[\mbox{curl}\,{\bf h}\times {\bf h}]_i%
\frac{\partial x_i}{\partial a_k}  \label{motion3}
\end{equation}
With the help of relation (\ref{htr}) and Eq. (\ref{motion}) the vector $%
{\bf u}$ given by (\ref{u}) will satisfy the equation
\begin{equation}
\frac{d{\bf u}}{dt}=\nabla \left( \frac{{\bf v}^2}2-w\right) -\frac 1{4\pi
}\left[ {\bf B}_0({\bf a})\times \mbox{curl}_a{\bf H}\right] .  \label{umhd}
\end{equation}
Here vector ${\bf B}_0({\bf a})={\bf h}_0({\bf a})/\rho_0({\bf a})$
is a Lagrangian invariant and ${\bf H}$
represents the co-adjoint transformation of the magnetic field, analogous to
(\ref{u}):
\[
H_i(a,t)=\frac{\partial x_m}{\partial a_i}h_m(x,t).
\]
Now by analogy with (\ref{motion}) and (\ref{weber}),
integration of Eq.(\ref{umhd}) over time
leads to the Weber type transformation:
\begin{equation}
{\bf u(a,}t)={\bf u}_0{\bf (a})+\nabla _a\Phi +\left[ {\bf B}_0({\bf a}%
)\times \mbox{curl}_a\tilde {{\bf S}}\right] .  \label{webermhd}
\end{equation}
Here ${\bf u}_0{\bf (a})$ is a new Lagrangian invariant which can be chosen
as pure transverse, namely, with $\mbox{div}_a$ ${\bf u}_0=0.$ This new
Lagrangian invariant cannot be expressed through the observed physical
quantities such as magnetic field, velocity and density. In spite of this
fact, as it will be shown in the next section, the vector Lagrangian
invariant ${\bf u}_0{\bf (a})$ has a clear physical meaning. As far as new
variables $\Phi $ and $\tilde {{\bf S}}$, they obey the equations:
\begin{eqnarray*}
\frac{d\Phi }{dt} &=&\frac{{\bf v}^2}2-w, \\
\frac{d\tilde {{\bf S}}}{dt} &=&-\frac{{\bf H}}{4\pi }+\nabla _a\psi .
\end{eqnarray*}
The transformation (\ref{webermhd}) for velocity ${\bf v(x},t)$ takes the
form:
\begin{equation}
{\bf v=}u_{0k}{\bf (a})\nabla a_k+\nabla \Phi +\left[ \frac{{\bf h}}\rho
\times \mbox{curl}\,{\bf S}\right]   \label{ZK}
\end{equation}
where ${\bf S}$ is the vector $\tilde {{\bf S}}$ transformed by means of the
rule (\ref{u}):
\[
S_i(x,t)=\frac{\partial a_k}{\partial x_i}\,\tilde S_k(a,t).
\]
In Eulerian description $\Phi $ satisfies the Bernoulli equation
\begin{equation}
\frac{\partial \Phi }{\partial t}+({\bf v\nabla })\Phi -\frac{{\bf v}^2}2+w=0
\label{bern}
\end{equation}
and equation of motion for ${\bf S}$ is of the form:
\begin{equation}
\frac{\partial {\bf S}}{\partial t}+\frac{{\bf h}}{4\pi }{\bf -}\left[ {\bf v%
}\times \mbox{curl}{\bf S}\right] +\nabla \psi _1=0.  \label{S}
\end{equation}
For ${\bf u}_0=0$ the transformation (\ref{ZK}) was introduced for ideal MHD
by Zakharov and Kuznetsov in 1970 \cite{ZK70}. In this case magnetic field $%
{\bf h}$ and vector ${\bf S}$ as well as $\Phi $ and $\rho $ are two pairs
of canonically conjugated variables. It is interesting to note that
in the canonical case the equations
of motion for ${\bf S}$ and $\Phi $ obtained
in \cite{ZK70} coincide with (\ref{bern}) and (\ref{S}).
However, the canonical parametrization describes
partial type of flows, in particular, it does not describe topological
nontrivial flows for which mutual knottiness between magnetic and vortex
lines is not equal to zero. This topological characteristics is given by the
integral $\int ({\bf v,h})d{\bf x.}$ Only when ${\bf u}_0\neq 0$ this
integral takes non-zero values.

\setcounter{equation} {0}
\section{Frozen-in MHD fields}

To clarify meaning of new Lagrangian invariant ${\bf u_0(a)}$ we
remind that the MHD equations (\ref{rho-t}-\ref{h-t}) can be obtained from
two-fluid system where electrons and ions are considered as two separate
fluids interacting each other by means of self-consistent electromagnetic
field. The MHD equations follow from two-fluid equations in the
low-frequency limit when characteristic frequencies are less than ion
gyro-frequency. The latter assumes i) neglecting by electron inertia, ii)
smallness of electric field with respect to magnetic field, and iii) charge
quasi-neutrality. We write down at first some intermediate system called
often as MHD with dispersion \cite{karpman}:
\begin{equation}
\mbox{curl}\ \mbox{curl}{\bf A}=\frac{4\pi e}c(n_1{\bf v}_1-n_2{\bf v}_2),
\label{varA}
\end{equation}
\begin{equation}
(\partial _t+{\bf v}_1\nabla )m{\bf v}_1=\frac ec\left( -{\bf A}_t+[{\bf v}%
_1\times \mbox{curl}\ {\bf A}]\right) -\nabla \frac{\partial \varepsilon }{%
\partial n_1},  \label{var1}
\end{equation}
\begin{equation}
0=-\frac ec\left( -{\bf A}_t+[{\bf v}_2\times \mbox{curl}\ {\bf A}]\right)
-\nabla \frac{\partial \varepsilon }{\partial n_2}.  \label{var2}
\end{equation}
In these equations ${\bf A}$ is the vector potential so that the
 magnetic field ${\bf h}=\mbox{curl}{\bf A}$ and electric field
${\bf E=-\frac 1cA_t}$. This system is closed by two continuity equations
for ion density $n_1$ and electron density $n_2$:
\begin{equation}
n_{1,t}+\nabla (n_1{\bf v}_1)=0,\qquad n_{2,t}+\nabla (n_2{\bf v}_2)=0.
\label{nepr2}
\end{equation}
In this system ${\bf v}_{1,2}$ are velocities of ion and
electron fluids, respectively. The first equation of this system
is a Maxwell equation for magnetic field
in static limit. The second equation is equation of motion for ions. The
next one is equation of motion for electrons in which we neglect by electron
inertia. By means of the latter equation one can obtain the equation of
frozenness of magnetic field into electron fluid (this is another Maxwell
equation):
\[
{\bf h}_t=\mbox{curl}[{\bf v}_2\times {\bf h}].
\]
Applying the operator $\mbox{div}$ to (\ref{varA}) gives with account of
continuity equations the quasi-neutrality condition: $n_1=n_2=n$. Next, by
excluding $n_2$ and ${\bf v}_2$ we have finally the MHD equations with
dispersion in its standard form \cite{karpman}:
\[
(\partial _t+{\bf v}\nabla )m{\bf v}=-\nabla w(n)+\frac 1{4\pi n}[\mbox{curl}%
\ {\bf h}\times {\bf h}],\qquad w(n)=\frac \partial {\partial n}\varepsilon
(n,n),
\]
\begin{equation}
n_t+\nabla (n{\bf v})=0,\qquad {\bf h}_t=\mbox{curl}\left[ \left( {\bf v}%
-\frac c{4\pi en}\mbox{curl}\ {\bf h}\right) \times {\bf h}\right] ,
\label{MHD2}
\end{equation}
where ${\bf v}_1={\bf v}$, and $\varepsilon(n,n)$ is internal energy
density so that $w(n)$ is entalpy per one pair ion-electron.
The classical MHD follows from this system in the limit when the last term $%
c/(4\pi en)\mbox{curl}\,{\bf h}$ in equation (\ref{MHD2}) should be neglected
with respect to ${\bf v}$. At the same time, the vector potential ${\bf A}$
must be larger characteristic values of $(mc/e){\bf v}$ in order to
provide inertia and magnetic terms in Eq. (\ref{var1}) being of the same
order of magnitude. Both requirements are satisfied if
$\epsilon =c/(\omega_{pi}L)<<1$ where $L$ is a characteristic scale of magnetic
field variation and $\omega _{pi}=\sqrt{4\pi ne^2/m}$ is ion plasma frequency.

Unlike MHD equations (\ref{rho-t}-\ref{h-t}), the given system  has two
frozen-in fields. These are the field ${\bf \Omega }_2=-\frac e{mc}{\bf h}$
frozen into electron fluid and the field
$$
{\bf \Omega }_1=\mbox{curl}({\bf v}%
+\frac e{mc}{\bf A})={\bf \Omega }-{\bf \Omega }_2
$$
frozen into ion component:
\[
{\bf \Omega }_{1t}=\mbox{curl}\left[ {\bf v\times \Omega }_1\right] ,
\]
\[
{\bf \Omega }_{2t}=\mbox{curl}\left[ {\bf v}_2{\bf \times \Omega }_2\right]
\]
where
$$
{\bf v}_2={\bf v}-\frac c{4\pi en}\mbox{curl}\ {\bf h.}
$$
Hence for both fields one can construct two Cauchy invariants by the same rule
(\ref{Cauchy}) as for ideal hydrodynamics:
\begin{equation}
{\bf \Omega _{10}(}{\bf a)}=J_1({\bf \Omega _1(x},t)\nabla ){\bf a(x,}t)
\label{C1}
\end{equation}
where ${\bf a(x,}t)$ is inverse mapping to ${\bf x=x}_1{\bf (a},t)$ which is
solution of the equation ${\bf \dot x}{={\bf v(x,}}t);$
\begin{equation}
{\bf \Omega _{20}(}{\bf a}_2)=J_2({\bf \Omega _2(x},t)\nabla ){\bf a}_2{\bf (x,%
}t)  \label{C2}
\end{equation}
with ${\bf a}_2{\bf (x,}t)$ inverse to the mapping ${\bf x=x}_2{\bf (a}_2,t)$
and ${\bf \dot x=v_2(x,}t).$

In order to get the corresponding Weber transformation for MHD as a limit of
the system it is necessary to introduce two momenta for ion and electron
fluids:
\begin{eqnarray}
{\bf p}_1 &=&m{\bf {v}}+\frac ec{\bf {A}} \\
{\bf p}_2 &=&-\frac ec{\bf A}.
\end{eqnarray}
In these expressions the terms containing the vector potential are greater
sum of ${\bf p}_1$ and ${\bf p}_2$ in parameter $\epsilon $. For each
momentum in Lagrangian representation one can get equations, analogous to
(\ref{Lag}), (\ref{motion}):
\begin{eqnarray}
\frac{\partial x_k}{\partial a_{1i}}\frac{dp_{1k}}{dt}&=&
-p_{1k}\frac{\partial v_k}{\partial a_{1i}}+
\frac \partial {\partial a_{1i}}
\left(-\frac{\partial\varepsilon}{\partial{n_1}}
+\frac{e}{c}({\bf v\cdot A})+m\frac{v^2}2\right)\\
\frac{\partial x_k}{\partial a_{2i}}\frac{dp_{2k}}{dt}&=&
-p_{2k}\frac{\partial v_{2k}}{\partial a_{2i}}+
\frac \partial {\partial a_{2i}}
\left(-\frac{\partial\varepsilon}{\partial{n_2}}
-\frac{e}{c}({\bf v}_2\cdot{\bf A})\right).
\end{eqnarray}
By introducing the vector $\tilde{{\bf p}}$ for each type of fluids,
by the same rule as (\ref{u}),
\[
\tilde{p}_i=\frac{\partial x_k}{\partial a_i}p_k,
\]
after integration over time of equations of motion for $\tilde{{\bf p}}$
one can arrive at two Weber transformations for each momentum:
\begin{eqnarray}
{\bf p}_1 &=&\tilde p_{1i}(a_1)\nabla a_{1i}+\nabla \Phi _1,  \label{p1} \\
{\bf p}_2 &=&\tilde p_{2i}(a_2)\nabla a_{2i}+\nabla \Phi _2.  \label{p2}
\end{eqnarray}
In the limit $\epsilon \rightarrow 0$ the
markers ${\bf a_1}$ and ${\bf a}_2$ can
be put approximately equal. This means that their difference will be small:
\[
{\bf a}_2-{\bf a_1}={\bf d \, \sim }\epsilon .
\]
Besides, due to charge quasi-neutrality, Jacobians with respect to $a_1$
and $a_2$ must be equal each other
(here we put $n_{10}({\bf a_1})=n_{20}({\bf a}_2)=1$
without loss of generality):
\[
\mbox{det}||{\partial {\bf x}}/{\partial {\bf a_1}}||=
\mbox{det}||{\partial {\bf x}}
/{\partial {\bf a_2}}||.
\]
As a result,  the infinitesimal vector ${\bf d(a,}t)$
relative to the argument
${\bf a}$ occurs divergence
free: $\partial d_i/\partial a_i=0.$

Then, summing (\ref{p1}) and (\ref{p2}) and considering the limit $\epsilon
\rightarrow 0$, we obtain the Weber-type transformation coinciding
with (\ref{webermhd}):
\begin{equation}
{\bf u(a,}t)={\bf u}_0{\bf (a})+\nabla _a\Phi +\left[ {\bf B}_0({\bf a}%
)\times \mbox{curl}_a\,\tilde {{\bf S}}\right] ,  \label{sum}
\end{equation}
where vectors ${\bf u}_0{\bf (a})$ and $\tilde {{\bf S}}$ are expressed
through the Lagrangian invariants $\tilde{\bf p}_1{\bf (a})$ and
$\tilde{\bf p}_2{\bf (a})$
and displacement ${\bf d}$ between electron and ion by means of relations
\cite{ruban}:
\begin{eqnarray*}
{\bf u(a,}t)&=&\frac 1m(\tilde{\bf p}_1{\bf (a})+\tilde{\bf p}_2{\bf (a})), \\
{\bf d}&=&-\frac{mc}e\mbox{curl}_a\tilde {{\bf S}}.
\end{eqnarray*}
Important that  in (\ref{sum}) all terms
 are of the same order of magnitude (zero order relative to $%
\epsilon) .$ Curl of vectors $\tilde{\bf p}_1{\bf (a})$ and
$\tilde{\bf p}_2{\bf (a}_2)$
yield the corresponding Cauchy invariants (\ref{C1}) and (\ref{C2}).

\setcounter{equation} {0}
\section{Relabeling symmetry in MHD}

Now let us show how existence of new Lagrangian invariants
corresponds to the relabeling symmetry.

Consider the MHD Lagrangian \cite{ZK},
\[
{\cal L}_*=\int\left(\rho\frac{{\bf v}^2}{2}-\rho\tilde\varepsilon(\rho) -%
\frac{{\bf h}^2}{8\pi}\right)d{\bf r},
\]
where we neglect by contribution from electric field in comparison
with that from magnetic field. Here $\tilde\varepsilon(\rho)$  is specific
internal energy.

In terms of mapping ${\bf x(a},t)$ the Lagrangian $%
{\cal L}_*$ is rewritten as follows \cite{IP}:
\begin{equation}  \label{L*}
{\cal L}_*=\int\frac{{\bf \dot x}^2}{2}d^3{\bf a}- \int \tilde\varepsilon(J_{%
{\bf x}}^{-1}({\bf a}))d^3{\bf a} -\frac{1}{8\pi} \int\left(\frac{({\bf %
h_0(a)\nabla_a)x}}{J_{{\bf x}}({\bf a})}\right)^2 J_{{\bf x}}({\bf a})d^3%
{\bf a}.
\end{equation}
Here density and magnetic field are expressed by means of relations
$$
\rho={1}/{J_{\bf x}},\qquad
{\bf h}=({\bf h}_0({\bf a})\nabla_a){\bf x}/J_{\bf x},
$$
and
$$
J_x({\bf a},t)= \mbox{det}||\partial {\bf x}/\partial {\bf a}||
$$
is the Jacobian of mapping ${\bf x=x(a,}t)$ and initial density is put to
equal 1.
Notice, that
variation of the action with by the Lagrangian (\ref{L*})
relative  to ${\bf x(a)}$ gives the equation of motion
(\ref{motion3}) (or the equivalent equation for vector ${\bf u}$ (\ref{umhd})).

Due to the presence of magnetic field in the Lagrangian (\ref{L*}),
the relabeling symmetry, in comparison with ideal hydrodynamics, reduces.
If the first two terms in (\ref{L*})
are invariant with respect to all incompressible changes
${\bf a\rightarrow a(b)}$ with $J|_{b}=1$, invariance of the last term,
however, restricts the class of possible deformations
up to the following class
$$
({\bf h_0(a)\nabla_a)b=h_0(b)}.
$$
For infinitesimal transformations
$$
{\bf a\rightarrow\ a+\tau g(a)}
$$
where $\tau$ is a (small) group parameter  the vector ${\bf g}$
must satisfy two conditions:
\begin{equation}
\label{condition}
\mbox{div}_a{\bf g}=0, \,\,\, \mbox{curl}_a[{\bf g\times h_0}]=0.
\end{equation}
The first condition is the same as for ideal hydrodynamics,
the second one provides conservation of magnetic field frozenness.

The conservation laws generating by this symmetry, in accordance
with Noether theorem, can be obtained
by standard scheme from the Lagrangian (\ref{L*}). They are written
through the infinitesimal deformation ${\bf g(a)}$ as integral over
${\bf a}$:
\begin{equation}
\label{conserv}
I=\int({\bf u,g(a)}) d{\bf a}
\end{equation}
where the vector ${\bf u}$ is given by (\ref{u}).
Putting ${\bf g=h_0}$ from this (infinite) family of integrals one gets the
simplest one
$$
I_{ch}=\int {\bf (v,h)}d{\bf r}
$$
which represents a cross-helicity characterizing degree of mutual knottiness
of vortex and magnetic lines.

The conservation laws (\ref{conserv}) are compatible with the
Weber-type transformation. Really, substituting (\ref{webermhd}) into
(\ref{conserv}) and using (\ref{condition}) one leads to the relation
$$
\int({\bf u_0(a),g(a)}) d{\bf a}.
$$
Hence conservation of (\ref{conserv}) also follows. Note that if
one would not suppose an independence of ${\bf u_0}$ on $t$ then, due to
arbitrariness of ${\bf g(a)}$,  this could be considered
as independent verification of conservation of solenoidal field ${\bf u_0}$:
$$
\frac d{dt}{\bf u_0} =0.
$$

The MHD equations expressed in terms of Lagrangian variables become
Hamiltonian ones, as in usual mechanics,
for momentum ${\bf p=\dot{{\bf x}}}$ and coordinate ${\bf x}$. These
variables assign the canonical Poisson structure.

In the Eulerian representation the MHD equations can be written
also in the Hamiltonian form
\cite{MG}:
$$
\rho_t=\{\rho, H \},\,\,\,
{\bf v}_t=\{{\bf v}, H \},\,\,\,
{\bf h}_t=\{{\bf h}, H \},
$$
where noncanonical Poisson bracket $\{F,G\}$ is given by the
expression (\ref{compres_bracket}).
As for ideal hydrodynamics, this Poisson bracket occurs to be degenerated.
For example, the cross helicity $I_{ch}$ serves a Casimir for
the bracket (\ref{compres_bracket}).
The reason of the Poisson bracket degeneracy is the same as for
one-fluid hydrodynamics - it is connected with a relabeling symmetry of
Lagrangian markers.

For incompressible case the Poisson bracket (\ref{compres_bracket}) reduces
so that it can be  expressed only through
magnetic field ${\bf h}$ and vorticity ${\bf \Omega}$:
\begin{equation} \label{brack}
\{F,G\}= \int\left( \frac{{\bf h}}{\rho}\cdot \left(\left[ \mbox{curl}\frac{%
\delta F}{\delta{\bf h}}\times \mbox{curl}\frac{\delta G}{\delta{\bf \Omega}}
\right] -
\left[ \mbox{curl}\frac{\delta G}{\delta{\bf h}}\times
\mbox{curl}
\frac{\delta F}{\delta%
{\bf \Omega}}\right] \right)\right)d^3{\bf r}
\end{equation}
\[
+\int\left( {\bf\Omega }\left[
\mbox{curl}\frac{\delta F}{\delta
{\bf\Omega}}\times \mbox{curl}\frac{\delta G}{\delta{\bf \Omega}}\right]
\right)d^3 {\bf r}.
\]
This bracket remains also degenerated.

\setcounter{equation}{0}
\section{Variational principle for incompressible MHD}

By analogy with incompressible hydrodynamics, one can introduce
magnetic line representation:
\begin{equation}  \label{hR}
{\bf h}({\bf r},t)=\int\delta({\bf r}-{\bf R}({\bf a},t)) ({\bf h}_0({\bf a}%
)\nabla_{{\bf a}}){\bf R}({\bf a},t)d^3{\bf a}.
\end{equation}
For vorticity the analog of vortex line parametrization (\ref{OmegaR})
can be obtained, for instance,
as a limit $\epsilon\rightarrow 0$ of the corresponding
representations for the two-fluid system. Calculations give \cite{ruban}:
\begin{equation}  \label{OmegaR1}
{\bf \Omega}({\bf r},t)=\int\delta({\bf r}-{\bf R}({\bf a},t)) (({\bf \Omega}%
_0({\bf a})+ \mbox{curl}_{{\bf a}} [{\bf h}_0({\bf a})\times{\bf U}({\bf a}%
,t)]) \nabla_{{\bf a}}){\bf R}({\bf a},t)d^3{\bf a},
\end{equation}
Here the field ${\bf {U}}({\bf {a}},t)$ is not assumed solenoidal,
as well as the Jacobian of mapping
${\bf r=R(a,}t)$ is not equal to unity.

 From the corresponding limit of the two-fluid system to incompressible MHD
it is possible also to get the expression for Lagrangian
\begin{equation}  \label{LAGRANGIAN1}
L=\int d^3{\bf a} ([({\bf h}_0\nabla_{{\bf a}}){\bf R}\times ({\bf U}\nabla_{%
{\bf a}}){\bf R}]\cdot{\bf R}_t)+
\end{equation}
\[
+{1}/{3} \int d^3{\bf a} ([{\bf R}_t \times {\bf R}]\cdot ({\bf \Omega}%
_0\nabla_{{\bf a}}){\bf R}) -{\cal H}\{{\bf \Omega}\{{\bf R,U}\},{\bf h}\{%
{\bf R}\}\}.
\]
The Hamiltonian of the incompressible MHD
${\cal H}_{MHD}$  in terms of ${\bf {U}}({\bf {a}},t)$ and ${\bf R(a,}t)$
takes the form
\[
{\cal H}_{MHD}= \frac{1}{8\pi}\int\frac {(({\bf h}_0({\bf {a}})\nabla_{{\bf {%
a}}}){\bf R}({\bf {a}}))^2} {\mbox{det}||\partial{\bf {R}}/\partial{\bf {a}}%
||}d^3{\bf {a}}+
\]
\begin{equation}  \label{H_MHD}
+\frac{1}{8\pi}\int\int \frac{ (({\bf \Omega}({\bf {a}}_1)\nabla_1){\bf R}(%
{\bf {a}}_1)\cdot ({\bf \Omega}({\bf {a}}_2)\nabla_2){\bf R}({\bf {a}}_2))}
{|{\bf {R}}({\bf {a}}_1)-{\bf {R}}({\bf {a}}_2)|}d^3{\bf {a}}_1d^3{\bf {a}}%
_2,
\end{equation}
where we introduce the notation
\[
{\bf {\Omega}}({\bf {a}},t)={\bf \Omega}_0({\bf a})+ \mbox{curl}_{{\bf a}}[{\bf %
h}_0({\bf a})\times{\bf U}({\bf a},t)].
\]

Equations of motion for  ${\bf {U}}$ and ${\bf {R}}$ follow from the variational
principle for action with Lagrangian (\ref{LAGRANGIAN1}):
\begin{equation}  \label{var_U}
\left[({\bf h}_0\nabla_{{\bf a}}){\bf R}\times {\bf R}_t\right]\cdot ({%
\partial{\bf R}}/{\partial a_{\lambda}})= -{\delta {\cal H}}/{\delta
U_{\lambda}},
\end{equation}
\begin{equation}  \label{var_R}
[({\bf \Omega(a,}t) \nabla_{%
{\bf a}}){\bf R}\times {\bf R}_t] -[({\bf h}_0\nabla_{{\bf a}}){\bf R}\times
({\bf U}_t\nabla_{{\bf a}}){\bf R}] ={\delta {\cal H}}/{\delta{\bf R}}.
\end{equation}
These equations can be obtained also directly from the MHD system
(\ref{rho-t}-\ref{h-t}) by the same scheme
as it was done for ideal hydrodynamics.

Thus, we have variational principle
for the MHD-type equations for two solenoidal
vector fields. Their topological properties
are fixed by ${\bf \Omega}_0({\bf %
a})$ and ${\bf h}_0({\bf a})$. These quantities represent  Casimirs
for the initial Poisson bracket (\ref{brack}).
It is worth noting that
the obtained equations of motion have the gauge invariant form. This gauge
invariance is a remaining symmetry connected with relabeling
of Lagrangian markers of magnetic lines in two-dimensional manifold
which can be specified always locally.
Coordinates of this manifold enumerate magnetic lines. This
symmetry leads to conservation of volume of magnetic tubes including
infinitesimally small magnetic tubes, namely, magnetic lines.
This property explains why  the Jacobian of the mapping ${\bf r=R(a,}t)$
can be not equal identically to unity.

\section*{Acknowledgments}

Authors thank A.B.Shabat for useful discussion of the connection between
NLSE and equations (\ref{dOmega_dt}), that resulted in integrability
declaration for (\ref{hamilt_chi}). This work was supported by the Russian
Foundation of Basic Research under Grant no. 97-01-00093 and by
the Russian Program for Leading Scientific Schools (grant no.96-15-96093).
Partially the work of E.K. was supported by the Grant INTAS 96-0413,
and the work of V.R.
by the Grant of Landau Scholarship.

\end{document}